\documentclass[aps,prl,showpacs,nofootinbib,amsmath,superscriptaddress,preprintnumbers,reprint]{revtex4-2}
\usepackage{graphicx}
\usepackage{dcolumn}
\usepackage{amsmath,amssymb}
\usepackage{bbold}
\usepackage{color}
\usepackage[dvipsnames]{xcolor}
\definecolor{darkblue}{rgb}{0,0,0.6}
\usepackage[colorlinks,linkcolor=darkblue,citecolor=darkblue,urlcolor=darkblue]{hyperref}
\usepackage{soul}
\definecolor{mygreen}{rgb}{0.0,0.55,0.3}

\newcommand{\juk}[1]{{\color{Brown}#1}}
\renewcommand{\epsilon}{\varepsilon}

\begin{document}

\title{Emerging mesoscale flows and chaotic advection in dense active matter}

\author{Yann-Edwin Keta*}

\affiliation{Laboratoire Charles Coulomb (L2C), Université de Montpellier, CNRS, 34095 Montpellier, France}

\author{Juliane U. Klamser*}

\affiliation{Laboratoire Charles Coulomb (L2C), Université de Montpellier, CNRS, 34095 Montpellier, France}

\author{Robert L. Jack}

\affiliation{Yusuf Hamied Department of Chemistry, University of Cambridge, Lensfield Road, Cambridge CB2 1EW, United Kingdom}

\affiliation{Department of Applied Mathematics and Theoretical Physics, University of Cambridge, Wilberforce Road, Cambridge CB3 0WA, United Kingdom}

\author{Ludovic Berthier}

\affiliation{Laboratoire Charles Coulomb (L2C), Université de Montpellier, CNRS, 34095 Montpellier, France}

\affiliation{Yusuf Hamied Department of Chemistry, University of Cambridge, Lensfield Road, Cambridge CB2 1EW, United Kingdom}

\let\thefootnote\relax\footnotetext{*These authors contributed equally to the work}

\date{\today}

\begin{abstract}
  We study two models of overdamped self-propelled disks in two dimensions, with and without aligning interactions. Active mesoscale flows leading to chaotic advection emerge in both models in the homogeneous dense fluid away from dynamical arrest, forming streams and vortices reminiscent of multiscale flow patterns in turbulence. We show that the characteristics of these flows do not depend on the specific details of the active fluids, and result from the competition between crowding effects and persistent propulsions. Our results suggest that dense active fluids present a type of `active turbulence' distinct from collective flows reported in other types of active systems.
\end{abstract}

\maketitle

Active matter has emerged as an important class of nonequilibrium systems, in which the injection of energy at the level of individual particles can produce emerging collective phenomena at large scale\juk{s}~\cite{marchetti2013hydrodynamics}. Among these, collective motion~\cite{vicsek2012collective} has attracted much interest because of its biological and social interest, \textit{e.g.} for wound healing~\cite{basan2013alignment} or crowd management~\cite{bain2019dynamic}.  Collective motion can be ordered, as in flocking~\cite{vicsek1995novel,toner1995longrange} where local interactions between individuals can lead to global motion along a given direction, or be more irregular or even chaotic, as in bacterial swarms~\cite{ariel2015swarming} or active nematics~\cite{doostmohammadi2018active} which display intermittent swirling motion.

The term `active turbulence'~\cite{alert2022active} recently became popular to describe chaotic mesoscale flows in various systems, from dense epithelial tissues~\cite{lin2021energetics} to suspensions of microtubules and kinesin motors~\cite{lemma2019statistical}. Unlike classical turbulence, active turbulence is observable in the absence of inertia. Moreover, the energy injection is not externally imposed but self-generated at small scale\juk{s}~\cite{alert2022active}. A recent classification~\cite{alert2022active} organises active turbulent models into four classes, depending on their symmetries:  A model's order parameter can be either polar or nematic; in addition, it is called ``wet'' if it conserves momentum, for example if hydrodynamic interactions dominate, and ``dry'' if it does not. 

In nematic systems~\cite{NemTurbGiomi2015,bar2020selfpropelled,Alert2020UniversalActiveTurb} flow derives in wet and dry conditions from an instability in the dynamics of the nematic director field, with an emerging length scale determined by the balance between active and nematic stresses~\cite{Alert2020UniversalActiveTurb,NemTurbGiomi2015}. Long-range velocity correlations in these flows are universal~\cite{Alert2020UniversalActiveTurb}. Most studies of polar active turbulence have  either considered wet systems of swimmers~\cite{chatterjee2021inertia}, or the Toner-Tu-Swift-Hohenberg equation~\cite{wensink2012mesoscale,mukherjee2023intermittency}, which describes incompressible flows in dry systems.   In this latter description, the polarisation and the velocity are assumed to be aligned: this is appropriate in the absence of steric interactions. Diverse particle-based models have also been shown to display some type of active turbulence: extensions of the Vicsek model~\cite{grossmann2014vortex,GrossmannTubAlignSperic2015}, self-propelled rods~\cite{wensink2012mesoscale,NumericsOfWensink2012mesoscale,Chate2018selfpropelled,bar2020selfpropelled} and dumbbells~\cite{turbGlassyDumbbellsDasgupta2017}, microswimmers with hydrodynamic interactions~\cite{qi2022emergence,HydroRodTurbStark2022}. All these models comply with the existing classification~\cite{Alert2020UniversalActiveTurb}.

\begin{figure*}[t]
\includegraphics[width=\textwidth]{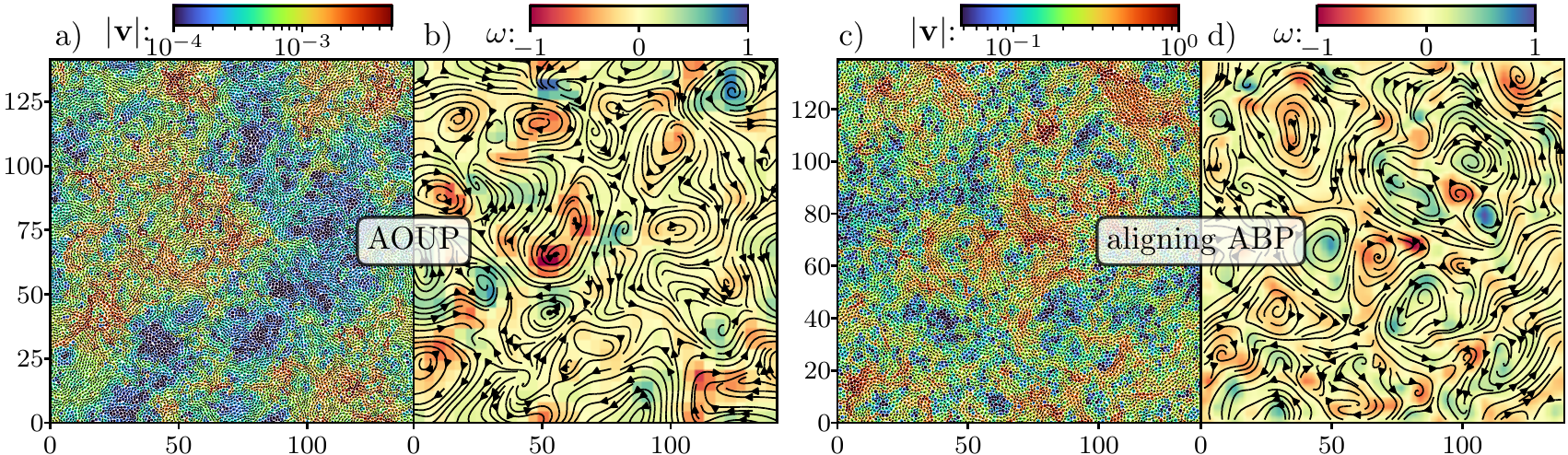}
\caption{\label{fig: Vorticity} (a) Configuration snapshot at $\phi = 0.8425$ of $N = 
16384$ AOUPs with velocity field (arrows) and corresponding velocity amplitude 
(color) showing fast and slow regions of collective motion for $\tau_p = 10^4$.
The corresponding vorticity field with streamlines in (b) highlights
the presence of streams and vortices in the velocity field. (c,d) are for $N =
12800$ aligning ABPs, that show a comparable phenomenology at $\phi = 0.97$ and
$\gamma = 2.5$.}
\end{figure*}

Here, we establish that the simplest class of active matter models -- overdamped self-propelled disks -- also develops mesoscale chaotic flows qualitatively similar to active turbulence, see Fig.~\ref{fig: Vorticity}. In two distinct models, we find that the homogeneous dense active fluid develops extended spatial velocity correlations~\cite{henkes2011active,szamel2015glassy,caprini2020active,caprini2020hidden,henkes2020dense,szamel2021longranged} which advect particles along a disordered array of streams and vortices, accompanied by hallmarks of active turbulence, including advective mixing. Within the existing symmetry classification~\cite{alert2022active}, the natural comparison is polar turbulence with dry friction~\cite{wensink2012mesoscale} but our results show different scaling behaviour. We attribute this to effects of particle crowding, which is absent from previous descriptions. Based on these observations we argue for a new class of active turbulent behaviour, which should encompass diverse models such as vibrated disks~\cite{deseigne2010collective}, self-aligning self-propelled particles~\cite{szabo2006phase,lam2015selfpropelled}, or self-propelled Voronoi models of confluent tissues~\cite{giavazzi2018flocking}.

We study $N$ overdamped athermal self-propelled particles in a square box of linear size $L$ with periodic boundary conditions that follow the overdamped dynamics
\begin{equation}
  \dot{\mathbf{r}}_i = - \mu \sum_{j \ne i}  \nabla_i U(r_{ij}) + \mu\mathbf{p}_i ,
\end{equation}
where $\mathbf{r}_i$ is the position of particle $i$, $\mathbf{p}_i$ the self-propulsion force, $\mu$ the particle mobility, $r_{ij} =  |\mathbf{r}_i-\mathbf{r}_j|$, and particles interact via a repulsive Weeks-Chandler-Andersen potential $U = 4\epsilon[(\sigma_{ij}/r_{ij})^{12} -  (\sigma_{ij}/r_{ij})^6 + 1/4]$ for $r_{ij} < 2^{1/6}\sigma_{ij}$ and $U = 0$ otherwise, where $\sigma_{ij} = (\sigma_i+\sigma_j)/2$ with  $\sigma_i$ the diameter of particle $i$.

The dynamics of the self-propulsion forces $\mathbf{p}_i$ defines the active model~\cite{fodor2016how}. We considered two distinct dynamics, active Ornstein-Uhlenbeck particles (AOUPs)~\cite{martin2021statistical,KetaJackBerthier2022} and aligning active Brownian particles (ABPs)~\cite{MartinGomez_AlignABP_2018,SeseSansa_AlignABP_2018,TailleurAlignABP2023,Tailleur_alignEffSteric2012}. To frustrate positional order, we introduce size polydispersity.   The diameters of the AOUPs are drawn from a uniform distribution of mean $\sigma = \overline{\sigma_i}$ and polydispersity 20\%~\cite{KetaJackBerthier2022,fily2014freezing}. The ABPs are a 50:50 bidisperse mixture with diameters $\sigma$ and $1.4 \sigma$. The packing fraction is $\phi = 2^{1/3} \pi N \overline{\sigma_i^2} / (4 L^2)$. The unit length is $\sigma$, the unit energy is $\epsilon$, and the unit time is $\mu \sigma^2 / \epsilon$. We measure velocities $\mathbf{v}_i = \dot{\mathbf{r}}_i - N^{-1}\sum_j \dot{\mathbf{r}}_j$ in the center-of-mass frame.

For AOUPs, the self-propulsion forces obey: 
\begin{equation}
\tau_p \dot{\mathbf{p}}_i = - \mathbf{p}_i + \sqrt{2D_0}\boldsymbol{\eta}_i ,
\label{Eq: AOUP}
\end{equation}
where $\tau_p$ is the persistence time, $D_0$ the diffusion constant of a free particle, and $\boldsymbol{\eta}_i$ is a Gaussian white noise of zero mean and unit variance, $\langle\boldsymbol{\eta}_i(t)\boldsymbol{\eta}_j(0)\rangle = \mathbb{1} \delta_{ij}\delta(t)$. From Eq.~(\ref{Eq: AOUP}), it follows that the amplitude of the self-propulsion force fluctuates around $\sqrt{ \langle | \mathbf{p}_i |^2 \rangle} = \sqrt{2 D_0/\tau_p}$. We use $D_0=1$, and vary $\tau_p$ towards large values.  We use system sizes up to $N=262144$ (depending on the state point), to ensure that results are not significantly affected by finite size effects.

For aligning ABPs, $\mathbf{p}_i = v_0 \mathbf{u}_i$ with a constant amplitude $v_0$ and orientations $\mathbf{u}_i = (\cos \theta_i, \sin \theta_i)$ evolving as
\begin{equation}
\dot{\theta}_i = \frac{\gamma}{n_i} \sum_j f(r_{ij}) \sin(\theta_j-\theta_i) + \sqrt{2D_r}\xi_i ,
\label{Eq: ABP}
\end{equation}
with $\gamma$ the alignment strength, $f(r_{ij}) = 1$ if $r_{ij}/\sigma_{ij} < 2$ and zero otherwise, $n_i = \sum_j f(r_{ij})$ the number of particles interacting with particle $i$, and $D_r$ the rotational diffusivity which controls the single-particle persistence time $\tau = D_r^{-1}$. We fix $v_0$ and $D_r$ to 1, and use modest $\gamma$ values, which are well below the onset of polar order. We use system sizes up to $N=51200$.

Fig.~\ref{fig: Vorticity} illustrates the emergent flows that are the main subject of this work (see \cite{movieLink} for corresponding movies): it displays velocity (${\bf v}$) and vorticity ($\nabla \times {\bf v}$, coarse-grained over a length 4) fields, as well as streamlines.  For suitable parameters, both models  support states where the density is homogeneous with clear signatures of active turbulence with non-trivial space and time fluctuations of the velocity field leading to mesoscale chaotic flows. The patterns in Fig.~\ref{fig: Vorticity} are highly dynamical and constantly form new networks of streams and vortices. Extended velocity correlations appear in a broad range of conditions (phase-separated~\cite{caprini2020spontaneous}, glassy~\cite{henkes2020dense,szamel2015glassy}, jammed~\cite{henkes2011active,ExtremeActiveMandal2020}, crystalline~\cite{caprini2020hidden}), but the active turbulent phenomenology discussed here is more complicated to observe.   

The emergence of active turbulence in AOUPs is surprising because there are no interactions favouring alignment of the self-propulsion forces, neither explicitly nor via shape anisotropy.  Instead, the flows emerge because extended velocity correlations are produced by the coupling between persistent self-propulsion and density fluctuations~\cite{caprini2020hidden,henkes2020dense,szamel2021longranged}. The relevant densities are large enough to avoid motility induced phase separation~\cite{TailleurMIPS2015} and small enough to avoid dynamic arrest~\cite{KetaJackBerthier2022}. For AOUPs under these conditions, advective flows develop gradually as $\tau_p$ increases~\cite{szamel2021longranged} [$\tau_p = 10^4$ in Figs.~\ref{fig: Vorticity}(a,b)]. This observation motivates our second model with weak alignment, in which similarly persistent self-propulsion arises from the aligning interactions, even if isolated particles decorrelate quickly ($\tau=1$). This drives aligning ABPs towards the same turbulent behaviour as highly-persistent AOUPs.

Despite differences in microscopic details, Fig.~\ref{fig: Vorticity} shows that the velocity correlations are almost indistinguishable in both models, as confirmed below. These similarities  support our identification of a new class of active turbulent systems, whose origin is the interplay of self-propulsion and crowding. In all cases, velocity correlations are much longer-ranged than the correlations of the self-propulsion forces $\mathbf{p}_i$ which are either absent (AOUPs) or weak (aligning APBs): velocity correlations are an emerging property. This situation is in contrast to the mechanism of correlated propulsions described by existing continuum theories~\cite{wensink2012mesoscale}, and supports our claim that these observations are not included in the current classification of active turbulent systems~\cite{alert2022active}.

\begin{figure}[t]
\includegraphics{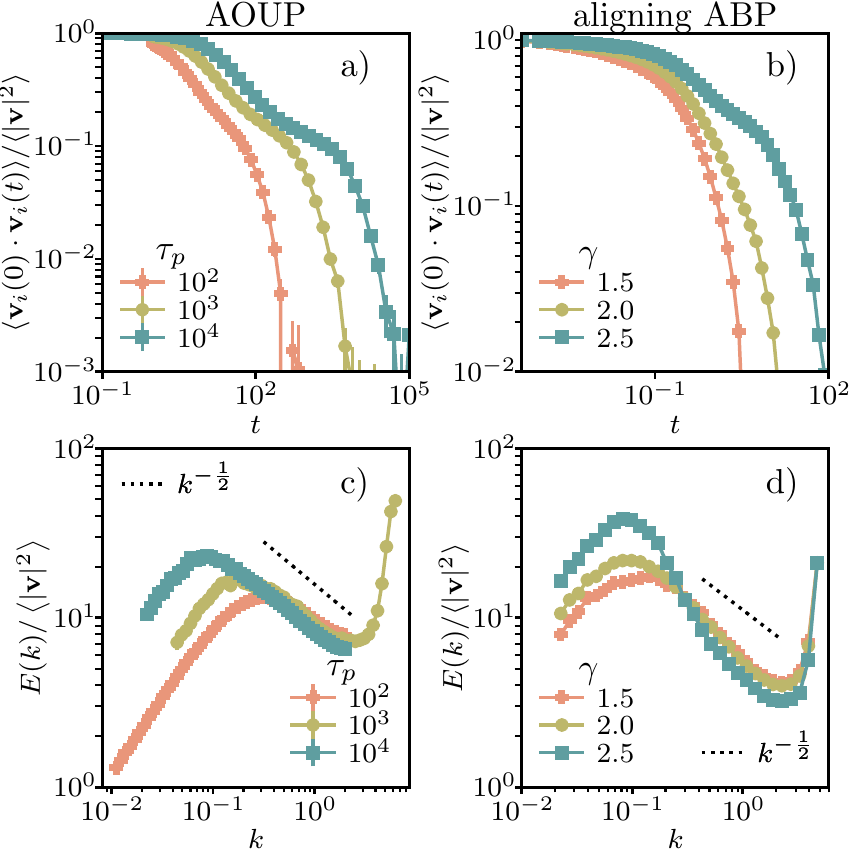}
\caption{\label{fig: < v_0 v_t > E(k)} (a,b) Velocity autocorrelations in time and (c,d)
kinetic energy spectra defined in Eq.~\eqref{eq:ek} for (a,c) AOUPs at 
various persistence times $\tau_p$ and (b,d) aligning ABPs for a range of 
alignment strengths $\gamma$.  For AOUPs, $\phi=0.84$ for $\tau_p=10^2,10^3$ and 
$\phi=0.8425$ for $\tau_p=10^4$. 
For ABPs, 
$\phi=0.97$.}
\end{figure}

\begin{figure}[t]
\includegraphics{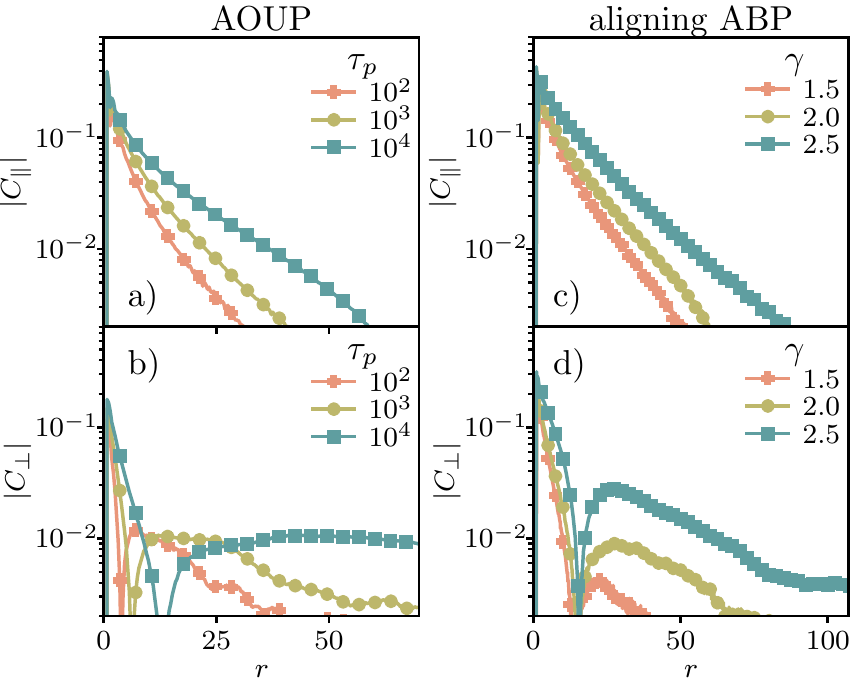}
\caption{\label{fig: C(r)} Real-space velocity correlations $C_\parallel(r)$ and 
$C_\perp(r)$ defined in Eq.~(\ref{Eq: C(r)}), for AOUPs and ABPs, as shown.
persistence times $\tau_p$ and alignment strength $\gamma$ respectively as 
indicated in the legend. 
The correlation length in $C_\parallel(r)$ (a,c) and 
the amplitude of negative correlations in $C_\perp(r)$ (b,d) can be tuned by 
increasing $\tau_p$ or $\gamma$ respectively.
Volume fractions $\phi$ are as in Fig.~\ref{fig: < v_0 v_t > E(k)}.}
\end{figure}

We now provide quantitative measurements supporting these conclusions. Figs.~\ref{fig: < v_0 v_t > E(k)}(a,b) show velocity autocorrelation functions, $\left< \mathbf{v}_i(0) \cdot \mathbf{v}_i(t) \right> / \left< | \mathbf{v} |^2 \right>$, which reveal the temporal behaviour of the flows. Unlike the exponential decay of simple fluids~\cite{hansen2007theory}, we observe a two-step decay in both models becoming more pronounced with more turbulent flows. These two time scales respectively correspond to the short collision time, and the increasing decorrelation time of the self-propulsion forces. 
In AOUPs, this longer correlation time  corresponds to the imposed persistence time $\tau_p$; in ABPs, it is controlled by the alignment strength $\gamma$ (recall that $\tau=1$ throughout).
  
We quantify spatial velocity correlations using the analog of the kinetic energy spectrum~\cite{wensink2012mesoscale}
\begin{equation} 
E(k) = \frac{2 \pi}{L^2} k  \left< \left| \tilde{\mathbf{v}}(\mathbf{k}) 
\right|^2 \right> , 
\label{eq:ek}
\end{equation} 
with $k = |\mathbf{k}|$ and $\tilde{\mathbf{v}}(\mathbf{k}) = \int  \mathrm{d}^2\mathbf{r} \, \mathbf{v}(\mathbf{r}) \exp{(-\mathrm{i}\mathbf{k}\cdot\mathbf{r})}$ the Fourier transform of the velocity field $\mathbf{v}(\mathbf{r}) = \sum_i  \mathbf{v}_i \delta(\mathbf{r}-\mathbf{r}_i)$, see Figs.~\ref{fig: < v_0 v_t > E(k)}(c,d). Clearly, $E(k)$ is directly related to velocity correlations in real space. For all parameters, $E(k) \sim k$ for small enough $k$, which implies the existence of a maximum length scale $\xi$ beyond which velocities are uncorrelated, so that $\langle| \tilde{\mathbf{v}}(\mathbf{k}) |^2\rangle = \mathrm{const}$ for $k \xi \ll 1$. This $\xi$ is the correlation length of the velocities.

For wave vectors $k$ intermediate between  $2\pi/\xi$ and $2 \pi/\sigma$, we report a decay of the energy spectrum $E(k) \propto k^{-\alpha}$ with $\alpha \simeq 1/2$. This corresponds to a scale-free decay $\sim r^{\alpha-1}$ of velocity correlations for length scales between the particle size $\sigma$ and the correlation length $\xi$~\cite{fisher1964correlation}. The established classes~\cite{alert2022active} of active turbulent behaviour involve significantly larger exponents (for example $\alpha=8/3$~\cite{wensink2012mesoscale}). Physically, $\alpha$ quantifies the observation that the velocity fields in Figs.~\ref{fig: Vorticity}(a,c) display self-similar structure up to the (parameter-dependent) correlation length $\xi$. For systems of non-aligning self-propelled particles, previous studies~\cite{szamel2021longranged,marconi2021hydrodynamics,henkes2020dense} reported results qualitatively similar to  those of Fig.~\ref{fig: < v_0 v_t > E(k)} but suggested a value $\alpha=1$, consistent with hydrodynamic models of self-propulsion coupled to density fluctuations. 

\begin{figure*}[t]
\includegraphics[width=\textwidth]{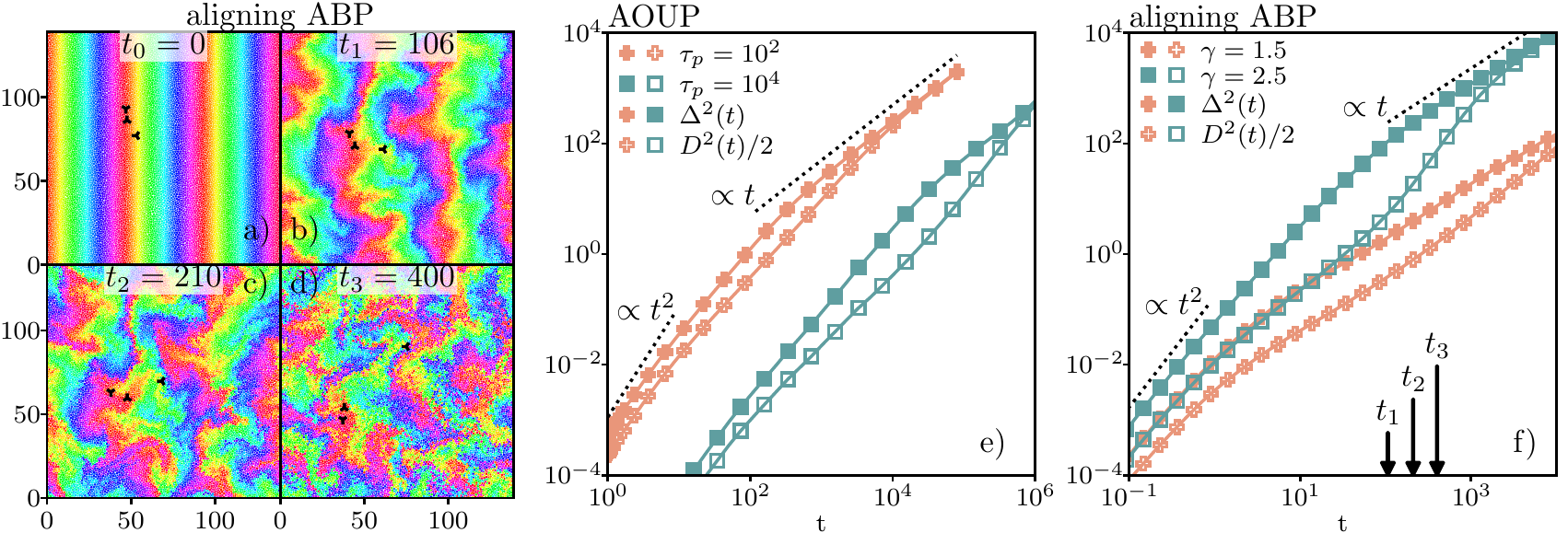}
\caption{\label{fig: Rainbow MSD}(a-d) Time series of configurations for aligning ABPs at $\gamma = 2.5$, $\phi = 0.97$. 
Particles are coloured according to their $x$ position at some time in the steady state denoted $t_0 = 0$. 
(e,f) Mean-squared displacement $\Delta^2(t)$ (full 
symbols) and mean-squared displacement difference of initially close by particles 
$D^2(t)$ (open symbols) for (e) AOUPs and (f) aligning ABPs.  The indicated times in (f) correspond to the snapshots in (b-d).
Volume fractions $\phi$ are as in Fig.~\ref{fig: < v_0 v_t > E(k)}.
}
\end{figure*}

To further understand and characterise these flow patterns, we decompose the real-space velocity correlations into longitudinal ($\alpha=\parallel$) and transverse ($\alpha = \perp$) components:  
\begin{equation}
C_\alpha(r) = \frac{\left\langle\sum_{i,j}   v_i^\alpha v_j^\alpha  \delta(r_{ij}-r) \right\rangle}{\left\langle\sum_{i,j} \delta(r_{ij}-r)\right\rangle} ,
\label{Eq: C(r)}
\end{equation}
where $v_i^\alpha$ is the velocity component in the direction parallel or transverse to the unit vector $(\mathbf{r}_i - \mathbf{r}_j) / r_{ij}$. The total velocity correlation function is $C(r)=C_\parallel(r)+C_\perp(r)$, but this decomposition is distinct from the Fourier analysis of~\cite{szamel2021longranged,henkes2020dense}, where $\textbf{v}$ is instead resolved parallel and perpendicular to the wave vector $\textbf{k}$. Fig.~\ref{fig: C(r)} shows results in both models, for a range of state points. The decomposition separates the long-ranged positive correlations along streams [in $C_\parallel(r)$], and the anti-correlations characteristic of vortices [in $C_\perp(r)$]~\cite{silverberg2013collective}. The data confirm a similar structure for both models, and show quantitatively that velocities are correlated over tens of particle diameters for the more persistent systems, in agreement with the peak position in $E(k)$ and the snapshots in Fig.~\ref{fig: Vorticity}. The characteristic size $\xi$ of the velocity patterns can be tuned via the persistence time $\tau_p$ of AOUPs, or the alignment strength $\gamma$ of aligning ABPs.  This leads in both cases to more extended streams and vortices. 

These emerging velocity correlations dramatically impact particle transport. This is revealed in Fig.~\ref{fig: Rainbow MSD} by `dyeing' particles according to their position at some initial time $t_0$ in the steady state, and watching them spread over time. Transport is dominated at initial times by rapid advection along extended streams, as revealed by the initial distortion of the pattern with mutually invading branches that stretch and fold over a range of length scales, resembling chaotic advection (see times $t_1$ and $t_2$). Only at large times do particles diffuse into regions of different colours which eventually blends the dyes. We also highlight three tracer particles which are initially very close, showing that particle pairs can be either advected large distances together or be separated almost immediately. These time-dependent patterns are qualitatively similar to the chaotic advection created for instance by time periodic flows~\cite{ottino1990mixing}.     

We quantify these observations using the mean-squared displacement $\Delta^2(t) = \langle | \Delta \mathbf{r}_i(t) |^2 \rangle$ and the mean-squared distance between initially close-by particles (as studied in inertial turbulence~\cite{JullienTurbSqSeparation1999,BourgoinTurbSqSep2015,OuelletteTurbMSD2011}), $D^2(t) = \langle | \Delta \mathbf{r}_i(t) - \Delta \mathbf{r}_j(t) | ^2\rangle$, where $\Delta\mathbf{r}_i(t) = \mathbf{r}_i(t) - \mathbf{r}_i(0)$ and the average is restricted to nearby pairs of particles with $|\mathbf{r}_i(0) - \mathbf{r}_j(0)| < 1.15 \sigma_{ij}$~\cite{shiba2012relationshipa}. 
By construction, both quantities vanish at $t=0$, while $D^2 \sim 2 \Delta^2 \sim t$ holds in the diffusive regime at large times (for which particles $i,j$ eventually decorrelate), see Fig.~\ref{fig: Rainbow MSD}(e,f).  

Self-propulsion causes ballistic motion $\Delta^2 \sim t^2$ at small times. The corresponding velocity decreases significantly for AOUPs as $\tau_p$ is increased at constant $D_0$, mirroring the reduction in strength of ${\bf p}_i$. In contrast, the velocity increases slightly with $\gamma$ for ABPs. This ballistic regime is quickly interrupted by interparticle collisions at a corresponding very small length scale. 
At very large times, memory of the self-propulsion forces is lost and particles diffuse, $\Delta^2 \sim t$. 
Between these two limits, we observe an intermediate advective (super-diffusive) regime, 
which is demarcated by the two well-separated time scales found in the velocity auto-correlation function (recall Fig.~\ref{fig: < v_0 v_t > E(k)}).  

The advection is also apparent in $D^2$ which is similarly ballistic at very short times. At intermediate times, $D^2$ grows significantly slower than $\Delta^2$ showing that pairs of particles can be  advected together over extremely large distances, leading to $D^2 \ll \Delta^2$.  Eventually, particles' memory of their initial conditions is lost: this leads to super-diffusive scaling, as $D^2$ `catches up' with the long-time diffusive scaling $D^2 \sim 2 \Delta^2 \sim t$.

In conclusion, we have established that a novel form of active turbulence generically emerges in two well-studied models of dry, isotropic, self-propelled particles. The observed mesoscale flows should be observable in a broad range of systems; they resemble other active chaotic flows, displaying scale-free behaviour from the particle size up to a correlation length scale that is easily tuned by the model parameters. However,  these flows  emerge here under the competition of highly persistent forcing and crowding in an otherwise homogeneous dense fluid. As previously developed theoretical descriptions of active turbulence rely on either polar or nematic interactions~\cite{alert2022active}, new approaches are needed that take into consideration the effect of steric crowding. Unusual transport properties emerge from the correlated velocity fields, including chaotic advection over large distances, which directly impacts mixing dynamics. Such properties may be useful when energy sources for the active particles are localised~\cite{Reichhardt2022ActiveResourceLandscape}, or in active matter with open boundaries~\cite{basan2013alignment}, or for mixtures of active particles~\cite{caballero2022activity}: all these cases deserve further studies. 

\acknowledgments
We thank D. Bartolo, J. Tailleur, and J. Yeomans for useful discussions. This work was publicly funded through ANR (the French National Research Agency) under the THEMA AAPG2020 grant. It was also supported by a grant from the Simons Foundation (\#454933, LB), and by a Visiting Professorship from the Leverhulme Trust (VP1-2019-029, LB). 

\bibliography{refs}

\end{document}